\documentclass{moriond}

\usepackage{amsmath}
\usepackage{booktabs}
\usepackage{color}

\def\be{\begin{equation}}
\def\ee{\end{equation}}
\def\bea{\begin{eqnarray}}
\def\eea{\end{eqnarray}}

\newcommand{\Photo}{}
\newcommand{\chbar}{$\overline{\mathrm{CH}}$}

\begin{document}

{\flushright{
        \begin{minipage}{5cm}
         DESY 15-084 \\
        \end{minipage}      }}
\vspace*{2cm}
\title{Estimation of uncertainties from missing higher orders in perturbative calculations}

\author{E.~Bagnaschi~$^{1}~$\footnote{emanuele.bagnaschi@desy.de}}

\address{
  {\sl $^{1}$DESY, Notkestra\ss e 85, D–22607 Hamburg, Germany}
}

\maketitle\abstracts{
  In this proceeding we present the results of our recent study~\cite{Bagnaschi:2014wea}~of the statistical performances of two different approaches,
  Scale Variation (SV) and the Bayesian model of Cacciari and Houdeau (CH)~\cite{Cacciari:2011ze} (which we also extend to observables with initial state hadrons),
  to the estimation of Missing Higher-Order Uncertainties (MHOUs)~\cite{David:2013gaa}, in perturbation theory.
  The behavior of the models is determined by analyzing, on a wide set of observables, how the MHOU intervals they produce are successful in predicting the next orders.
  We observe that the Bayesian model behaves consistently, producing intervals at $68\%$ Degree of Belief (DoB) comparable with the scale variation intervals with
  a rescaling factor $r$ larger than $2$ and closer to $4$. Concerning SV, our analysis allows the derivation of a heuristic Confidence Level (CL) for the intervals. 
  We find that assigning a CL of $68\%$ to the intervals obtained with the conventional choice of varying the scales within a factor of two
  with respect to the central scale could potentially lead to an underestimation of the uncertainties in the case of observables with initial state hadrons.
}

\section{Introduction}

Precision phenomenology will be of primary importance for new physics searches during the upcoming second run of the LHC, since no strong hints of deviation from the Standard Model (SM)
have emerged from the analysis of run-1 data. These analyses require accuracy not only in the experimental measurements (i.e.~high statistics and strict control of systematic errors)
but also in the SM theoretical predictions. When experimental and theoretical uncertainties are of the same order of magnitude, 
it becomes important to be able to assess quantitatively the importance of all sources of uncertainties.
Concerning theoretical uncertainties, one of the most important classes is represented by the uncomputed higher-order contributions to the perturbative expansion of the observables.
Traditionally, the community has estimated them by varying the unphysical scales that appear in the perturbative result (e.g.~the renormalization scale and the factorization scale).
This procedure has been used for many years and it allows a quick estimate of the effects of missing higher orders.
However this method yields intervals that do not posses a strict statistical meaning, therefore making it difficult to integrate these uncertainties in more complex analyses of the experimental data. 
To solve some of the shortcomings of the SV procedure, Cacciari and Houdeau have recently proposed a new method, built using the framework of Bayesian statistics~\cite{Cacciari:2011ze}.
While not considered in this proceeding, it should be pointed out that the same problem has also been studied with a more mathematical oriented approach, using sequence transformation techniques, in ref.~\cite{David:2013gaa},
where also the concept of MHOU was originally introduced.

In our study~\cite{Bagnaschi:2014wea}, we first try to address some of the drawbacks of the original CH framework, introducing a modified version, named \chbar, which we also extend to observables with initial state hadrons.
We then consider the two models, SV and \chbar, and we analyze their performances on two wide set of observables, characterized by the presence/absence of hadrons in the initial state.
This allows us to test the consistency of the \chbar~model, estimate a heuristic confidence level for the SV intervals and then compare the two prescriptions for the estimation of the MHOUs.

\section{Models for the estimation of theoretical uncertainties}

\subsection{Scale variation}
The first model we consider is the standard and widely used procedure of scale variation.
Given an arbitrary perturbative observable O, characterized by a central hard scale $Q$, its truncated perturbative expansion
\be
	O_k(Q,\mu) =\sum\limits_{n=l}^{k}\alpha_s^n(\mu) c_n(Q,\mu)
	\label{eq:pertexp}
\ee
contains a residual dependence on the unphysical scales (e.g.~the renormalization scale), which here we collectively represent with $\mu$, that were introduced during the computation.
This dependence is unphysical and will vanish if the observable were computed at all orders. Specifically, the functional dependence on $\mu$ of the expression 
given in eq.~\ref{eq:pertexp} is of higher order in $\alpha_s$ and it is governed by the renormalization group equation.
The latter is such that varying the scales yields terms that are effectively part of the unknown higher-order contribution,
multiplied by logarithmic factors of the ratio of the scales $Q$ and $\mu$. Due to these properties, it is customary to evaluate the degree of convergence
of a perturbative expansion (i.e.~the size of the missing higher-order terms) by varying the unphysical scales around their central value Q in the interval
$[Q/r,rQ]$, where the rescaling factor $r$ is usually chosen equal to $2$. Several prescriptions for SV are adopted in the literature,
differing in the details of how the uncertainty interval is constructed from the values obtained when the scales are varied (e.g.~scanning vs taking the values at the extremes of the variation interval).

However the SV method has two drawbacks. First of all, the intervals obtained in this way have a priori no statistical meaning and therefore can not be included consistently in other analyses.
Secondly, the value for the rescaling factor $r$, usually assumed to be two, is arbitrary and there are no clear-cut theoretical justifications to opt for a specific value over another.

\subsection{The Cacciari-Houdeau Bayesian framework}

As a second prescription, we consider the Bayesian framework developed by Cacciari and Houdeau~\cite{Cacciari:2011ze}, originally introduced to address the shortcomings of SV.
One of the drawbacks of the original CH model is that its prediction for the uncertainty intervals depends on the expansion parameter chosen for the perturbative series of the observable.
However, in perturbative QCD, the expansion parameter is not unambiguously defined. To address these issues, we introduce a parameter $\lambda$ that reflects our ignorance of the optimal expansion parameter.
Our expression for a generic observable, which starts at order l and it is known up to order k, is then given by
\be
	O_k = \sum\limits_{n=l}^{k}\left(\frac{\alpha_s}{\lambda}\right)^n (n-1)!~\lambda^n \frac{c_n}{(n-1)!} \equiv \sum\limits_{n=l}^{k}\left(\frac{\alpha_s}{\lambda}\right)^n (n-1)!~b_n,
	\label{eq:chbarexp}
\ee
where we have isolated a factor $(n-1)!$ in the expansion. The latter can be motivated from theory by looking at the behavior of the coefficients $c_n$ for large $n$. The same priors of the original CH model are retained. Then, we extrapolate $\lambda$ by measuring the performance of the model for a fixed $\lambda$ value on a given set of observables. 
At a given order and for a given DoB, we calculate for each observable the corresponding uncertainty interval. Then we compute the success rate, defined as the ratio between
the number of observables whose next-order is within the computed DoB interval over the total number of observables in the set. We define the optimal $\lambda$ value to be such that the success rate is equal to requested DoB. 

We then perform the extension to hadronic observables 
\be
 O_k (\tau,Q) = \mathcal{L}(Q) \otimes \sum_{n=l}^k \alpha_s^n C_n(Q)
 \label{eq:genhadr}
\ee
in two possible ways:
\begin{enumerate}
  \item By performing the convolution integral in eq.~\ref{eq:genhadr} (using the same PDFs at all orders), extracting the perturbative coefficients by re-expanding in power of $\alpha_s$ with the same factors as in the non-hadronic case
        and then by feeding them to the same model.
      \item By going to Mellin space, identifying the dominant~\footnote{We follow the observations, first appeared in ref.~\cite{Bonvini:2010tp}, on the analytic structure of the cross section in Mellin space and its dominant contributions.} Mellin moment $N^*$, applying the \chbar~model to the coefficient function at this moment and then rescaling back the uncertainty interval to the observable in the physical space. This procedure is theoretically cleaner,
        since it does not involve non-perturbative physics coming from the PDF.
\end{enumerate}
For our global analysis we opted for the first method, due to its simplicity and ease of use. However we have also applied the second approach to some observables, like $pp \to H$.
In all cases the value of $\lambda$ should be re-determined with the same procedure that was used for the observables without initial state hadrons. We denote the new value with $\lambda_h$.

\section{Results}
\begin{figure}
  \includegraphics[width=0.5\textwidth]{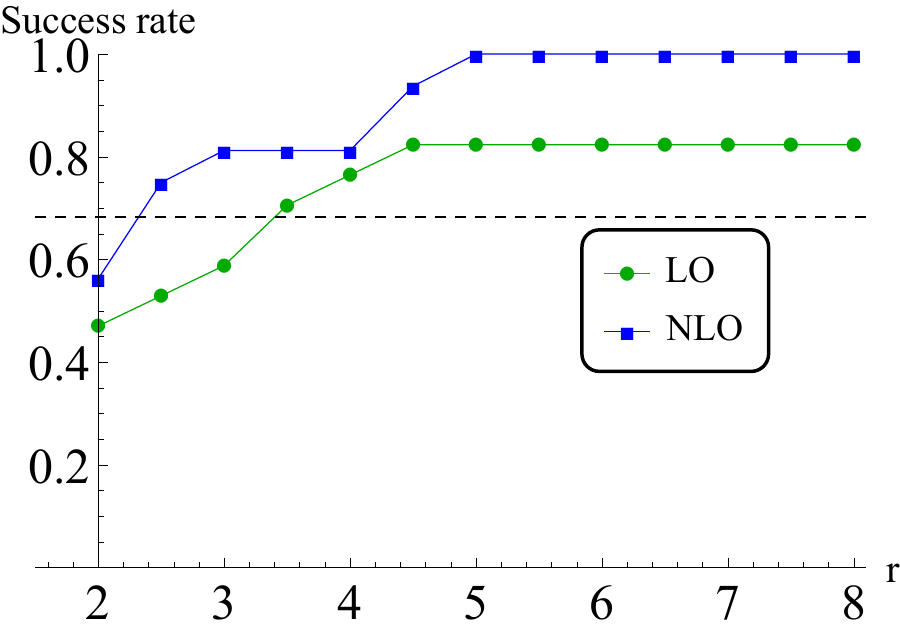}\hfill
  \includegraphics[width=0.5\textwidth]{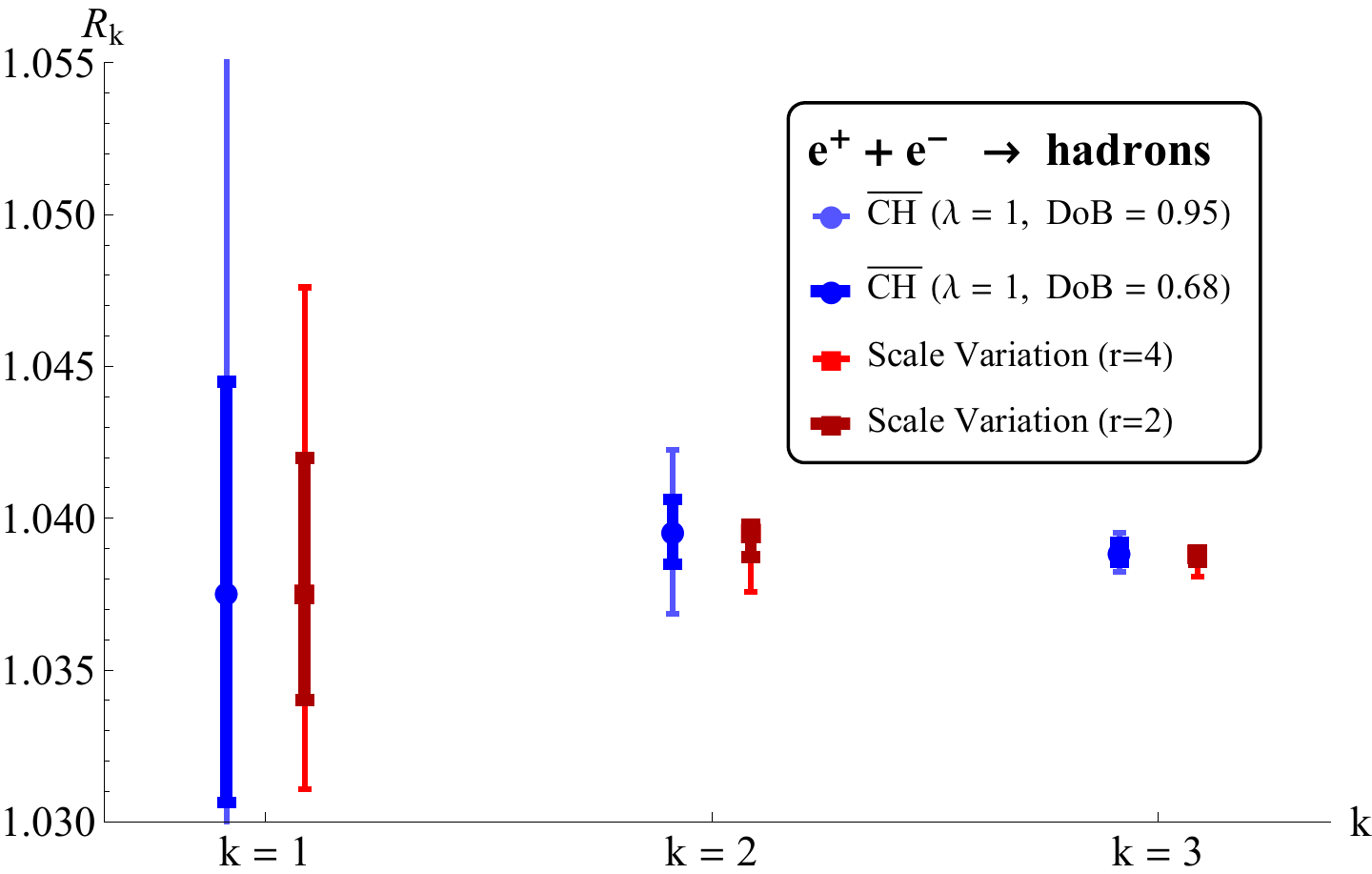}
  \caption{On the left we show the results for the SV heuristic CL, for the non-hadronic observable set, considering the values assumed by the observable in the entire scale variation interval.
           On the right we show the comparison of the error bars, for $e^+e^- \to \mathrm{hadrons}$, produced by SV (red) and the \chbar~model (blue) for different values of $r$ and DoB, respectively.}
  \label{fig:nonhadronic}
\end{figure}

We consider two sets of observables, characterized by the presence or absence of hadrons in the initial state. The former set includes the $R$ ratio, QCD sum rules, event shape observables, particle decays and splitting kernels.
The latter contains Higgs and weak bosons production (also in association with jets), $t\bar{t}$ and $b\bar{b}$ production.
For each observable in the set under study, we compute the MHOU at a given order $k$ with a specific model and we compare with the prediction at order $k+1$. We define the success rate of the model as the ratio of the observables for which the $k+1$ order is inside the error band over the total number of observables in the set. In the case of the \chbar~model, this procedure is repeated for different DoBs and it is also used to determine $\lambda$ at the same time, while for SV the algorithm is repeated for different values of $r$.

In the left plot of fig.~\ref{fig:nonhadronic}, which shows the performance of SV for the non-hadronic observables, we notice that at LO (as expected) SV is poorly predictive of the next unknown-order up to $r=3-4$,
while at NLO we observe a heuristic $68\%$ CL already for $r \simeq 2$. The comparison of the results for SV and \chbar~(with $\lambda = 1$, determined as explained above),
on the right side of the same figure, allows us to compare the predictions from the two models for a specific observable, $e^+e^- \to \mathrm{hadrons}$. We see that the $68\%$ DoB Bayesian intervals are slightly larger than the $r=2$ SV intervals and comparable in size with the $r=4$ ones.

\begin{figure}
  \includegraphics[width=0.5\textwidth]{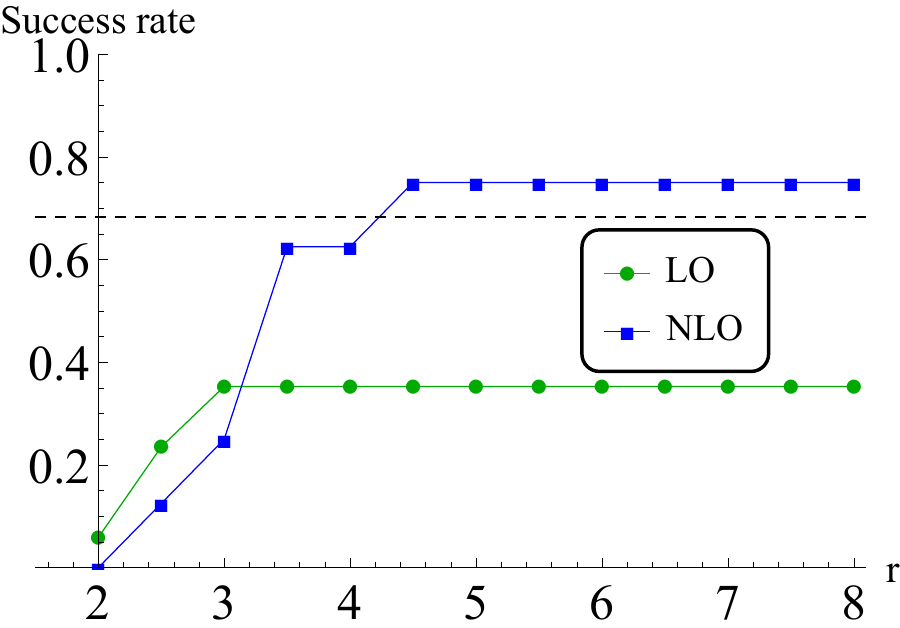}\hfill
  \includegraphics[width=0.5\textwidth]{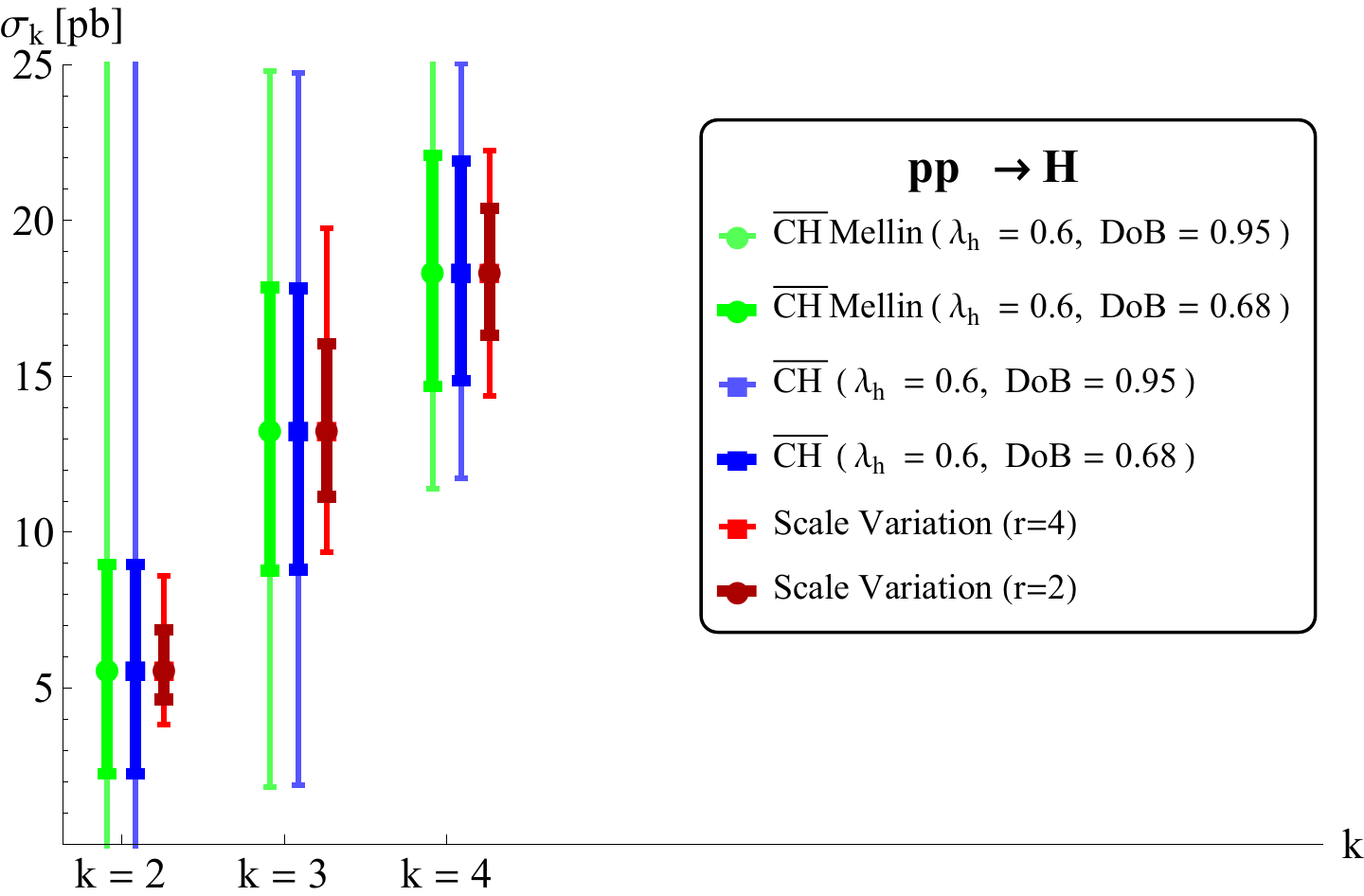}
  \caption{On the left we show the results for the SV heuristic CL for the hadronic observable set, as a function of $r$,
    with order-matched PDFs and by considering only the observable values at the extremes of the scale variation interval.
   On the right we show the comparison of the error bars, for $pp \to H$, produced by SV (red) and the \chbar~model (blue and green) for different values of $r$ and DoB, respectively.}
  \label{fig:hadronic}
\end{figure}

In fig.~\ref{fig:hadronic}, we show the same analysis of fig.~\ref{fig:nonhadronic} for the hadronic-observable set.
From the left plot, we see that the SV intervals are less predictive than in the non-hadronic case and, even at NLO, a heuristic CL of $68\%$ is obtained only for $r \simeq 4-5$, though the specific value
depends whether one uses the NNLO PDF at all orders, or, as it is done in the plot, matches the PDF to the perturbative order of the computation.
On the right, we show, as an example, the results for $pp \to H$. We see that, for this specific observable, the bands obtained with the \chbar~model~\footnote{In the hadronic case, we have found $\lambda_h \simeq 0.6$, reflecting systematically larger radiative corrections than in the non-hadronic one. Indeed smaller values of $\lambda$ correspond to larger error bars in the \chbar~model. } are systematically larger than the ones obtained with SV.
We also observe that the two possible extensions to hadronic observables of the \chbar~model agree in their prediction for the MHOU.

\section{Conclusions and outlook}

We have performed a statistical study of the performances of the SV and \chbar~models. Concerning SV, we find that, for non-hadronic observables, a value of $r \simeq 2$ corresponds to a heuristic
CL of about $68\%$ at NLO. On the other hand, in the hadronic case, a larger value of $r$, between $4$ and $5$, is needed to reach the same confidence level (the precise value being dependent on the choice adopted for the PDFs).
With respect to the \chbar~model, which we have also extended to hadronic observables, we find that a consistent determination of the value of $\lambda$ is possible
and that it leads to error bars that are of the same order of magnitude of the SV intervals, with a $68\%$ DoB band usually slightly larger than the $r=2$ and closer to the $r=4$ SV interval.
We have therefore shown that the \chbar~model can be reliably used to estimate MHOUs. Possible future improvements involve the introduction of a specific prior for the value of $\lambda$, to replace the current
frequentist procedure, and the development of more refined models tailored on specific classes of observables, especially for the hadronic case.

\end{document}